# A simple mechanism for the generation of Earth's magnetic field

Oleg. V. Styazhkin

**Abstract:** *Based on the ideal gas model, the dielectric polarization of mantle is achieved, the physical-mathematical model is constructed and the estimate calculation of a dipole of Earth's magnetic field with the considering of rotation, parameters of density and temperature, potential of ionization and static dielectric constant (relative dielectric permittivity), chemical compound of substance in Earth's mantle is executed.*

-----

# Einfacher Mechanismus zur Erzeugung des Erdmagnetfeldes

Oleg. V. Styazhkin

**Abstract:** *Aufgrund vom Modell des idealen Gases wird die elektrische Verschiebung freier Elektronen im Erdmantel bekommen, physisch-mathematisches Modell aufgrund der Erdrotation und Parametern der Dichte, Temperatur, Dielektrizitätszahl, Ionisierungsenergie und des Prozentsatzes der chemischen Hauptverbindungen des Erdmantels aufgebaut und die Bewertungsberechnungen des Dipolmoments erfüllt.*

**Stichwörter:** Das Modell des idealen Gases, Maxwell-Boltzmann-Verteilung, das Erdmagnetfeld, Dipolfeld, die physische Parameter des Erdmantels.

-----

О.В.Стяжкин

# ПРОСТОЙ МЕХАНИЗМ ГЕНЕРАЦИИ ГЕОМАГНИТНОГО ПОЛЯ

*Аннотация:* На базе модели идеального газа получена поляризация зарядов в мантии, построена физико-математическая модель и выполнены оценочные расчёты дипольной моды магнитного поля Земли с учётом скорости её углового вращения, параметров плотности, температуры, химического состава, потенциала ионизации, диэлектрической проницаемости и процентного содержания основных химических соединений вещества мантии.

*Ключевые слова:* модель идеального газа, статистика Максвелла-Больцмана, магнитное поле Земли, дипольная мода магнитного поля, физические параметры мантии Земли.

## ВВЕДЕНИЕ

В настоящее время достигнуты большие успехи в изучении механизма эволюции магнитного поля Земли на базе модели гидромагнитного динамо [8], действующего предположительно в жидком ядре Земли. Построены экспериментальная установка [11] и простая математическая модель [9], демонстрирующие процесс инверсии магнитного поля. Последняя работа демонстрирует возможность инверсии магнитного поля на фоне его стационарной дипольной составляющей, что позволяет рассматривать скорость вращения Земли в качестве одного из параметров механизма генерации магнитного поля. Существуют также работы, рассматривающие дополнительные механизмы генерации, например [12]. В данной работе на основе модели идеального газа предложен простой механизм генерации геомагнитного поля, действующий предположительно в мантии Земли. На базе физических параметров мантии и с учётом скорости вращения Земли получены результаты, демонстрирующие хорошее совпадение расчётных и измеренных значений магнитного момента.

## ОСНОВНЫЕ ФИЗИЧЕСКИЕ ПАРАМЕТРЫ ЗЕМЛИ

### Характеристика магнитного поля Земли

В первом приближении земной магнит представляется диполем, наклонённым к оси вращения под углом 11° и имеющим на магнитном экваторе напряженность 0,3 G. К настоящему времени измерены амплитуды более десятка следующих за диполем гармоник, которые уменьшаются по степенному закону с изломом на восьмой гармонике. На дипольную моду приходится около 90% напряженности. Остаточное поле (полное минус диполь) имеет вид конечного числа аномалий, занимающих области размерами от сотен до двух тысяч километров.

Наблюдаются хаотические флуктуации направления дипольного момента с характерными временами $10^3$-$10^4$ лет. При усреднении по этим флуктуациям средний земной диполь будет ориентирован вдоль оси вращения. Следовательно, вращение оказывает сильное влияние на эволюцию магнитного поля. За характерное время порядка $10^5$ лет происходят обращения (инверсии) направления магнитного диполя. Процесс случайный [8, с.266-267].

Моды выше дипольной рождаются из турбулентных флуктуаций течений электропроводящей жидкости во внешнем ядре Земли. Математическое моделирование показывает, что если их рассматривать как белый шум, действующий на дипольную моду, то возникают состояния модели, объясняющие инверсию магнитного поля Земли [9].

### Физические и химические параметры мантии Земли

На рис.1 представлена упрощенная схема строения Земли, где указаны названия областей и расстояния от поверхности до границ характерных состояний вещества [1, Т2, с.79].

| Внутреннее и Внешнее ядро | | Нижняя и Верхняя мантия | | Кора |
|---|---|---|---|---|
| 6371 *км* | 5120 *км* | 2885 *км* | 1000 *км* | 40 *км* |

**Рис.1.** Схема строения Земли.

Распределение по глубинам давления, температуры и плотности по модели «Земля-2» В. Н. Жаркова, В. П. Трубицына и П. В. Самсоненко приведено в табл.1, [3, с.26-27]. Выполнена интерполяция параболическими сплайнами табличных данных температуры и плотности вещества в недрах Земли.

Диапазон значений плотности (5.56-10.08) *г/см*$^3$ на глубине 2920 *км* заменён средним значением 7.82 *г/см*$^3$. Параметр глубины пересчитан на расстояние от центра. Параметр плотности нормирован к массе Земли. Все параметры приведены к международной системе измерений СИ.

Полученные зависимости абсолютной температуры $T_z(r)$ и плотности $\rho_n(r)$ вещества в недрах Земли, используемые в дальнейшем при расчете ионизации и поляризации свободных электронов в мантии Земли, приведены на рис.2. Вертикальными маркерами отмечены радиус внутреннего ядра $R_j = 1.25$ Mm и внутренний радиус мантии $R_{mu} = 3.45$ Mm. Наибольшее значение радиуса соответствует верхней границе мантии $R_{mo} = 6.34$ Mm.

**Табл.1.** Давление, плотность и температура в недрах Земли.

| Глубина, км | Давление, мегабары | Плотность, г/см³ | Температура, °C |
|---|---|---|---|
| 30 | 0.0084 | 3.32 | 700 |
| 100 | 0.031 | 3.38 | 1500 |
| 200 | 0.065 | 3.46 | 1950 |
| 413 | 0.130 | 3.64 | 2400 |
| 1047 | 0.399 | 4.58 | 2800 |
| 2060 | 0.889 | 5.12 | 3600 |
| 2920 | 1.386 | 5.56-10.08 | 4300 |
| 3955 | 2.445 | 11.46 | 5250 |
| 4991 | 3.239 | 12.28 | 6050 |
| 6371 | 3.657 | 12.68 | 6300 |

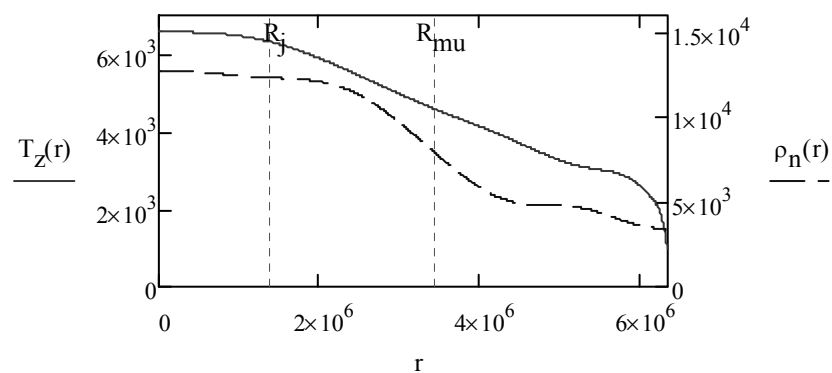

**Рис.2.** Температура $T_z(r)$ и плотность $p_n(r)$ в недрах Земли.

В табл.2 приведено содержание окислов основных элементов мантии Земли в массовых процентах (по А.Э. Рингвуду) [4, с.12], их молекулярный вес, количество атомов в молекуле, плотность, относительная диэлектрическая проницаемость [5], первый потенциал (энергия) ионизации молекул [6] и [7]. Окислы, процентное содержание которых меньше 1%, из расчета исключены.

**Табл.2.** Физические параметры окислов основных элементов мантии.

| Окисел | SiO₂ | MgO | FeO | Al₂O₃ | CaO |
|---|---|---|---|---|---|
| Массовая доля в% [4, с.12] | 45.1 | 38.1 | 7.6 | 4.6 | 3.1 |
| Молекулярный вес, в атомных единицах массы [5, Т.2] | 60.8 [с.104] | 40.31 [с.114] | 71.84 [с.60] | 101.96 [с.20] | 56.08 [с.92] |
| Количество атомов | 3 | 2 | 2 | 5 | 2 |
| Плотность, г/см³ [5, Т.2] | 2.3 [с.104] | 3.58 [с.114] | 5.7 [с.60] | 3.97 [с.20] | 3.37 [с.92] |
| Диэлектрическая проницаемость [5, Т.1, с.960] | 3.9 | 9.65 | 100 | 8.8 | - |
| Потенциал (энергия) ионизации, V (eV) | 11.7 [6] | 8.5 [6] | - | 9.9 [7] | 6.5 [6] |

Приведенные данные использованы в дальнейшем при расчете ионизации и поляризации зарядов в мантии Земли.

### Ионизация мантии Земли

Для нижней оценки степени ионизации мантии достаточно учёта первого ионизационного потенциала. Число молекул, имеющих энергии больше заданной $w_0$, при $w_0 \gg k_b T$, можно представить выражением [10, с.208]:

$$n(w_0) = 2n\sqrt{w_0 / \pi \cdot k_b T} \cdot e^{-w_0/k_b T}$$

где: $n$ – общее число молекул в рассматриваемом объёме, $k_b$ - постоянная Больцмана, $T$ - температура вещества. Это выражение справедливо и для плотности молекул, если взять производную по объёму от обеих частей, а число молекул в единице объёма является статистически значимым.

Так как вероятность ионизации находится в экспоненциальной зависимости от энергии, нижняя оценка плотности ионов в однородной смеси характеризуется плотностью вещества с наименьшей энергией ионизации:

$$\delta_i(r) = 2m_i(r)\sqrt{w_i / \pi \cdot k_b T_z(r)} \cdot e^{-w_i/k_b T_z(r)} \qquad (1)$$

где: $m_i(r) = \rho_n(r) \cdot n_i/m_u \cdot M_i$ - плотность молекул ***i***-го окисла в смеси, $\rho_n(r)$ - плотность вещества в недрах Земли, $n_i = pr_i/\Sigma\, pr_i$ - доля ***i***-го окисла в смеси, $pr_i$ - процентное содержание окисла (табл.2), $m_u$ - константа атомной единицы массы, $M_i$ - молярная масса молекулы $i$-го окисла (табл.2); $w_i$ - первый потенциал ионизации молекулы (табл.2), $T_z(r)$ - температура в недрах Земли.

Ниже приведен расчет ионизации в характерных точках для однородной смеси, в которой CaO обладает наименьшей энергией ионизации, табл.2. Логарифм относительной плотности ионов $\delta_i(r)$ в мантии Земли представлен на рис.3. Вертикальными маркерами отмечены условные границы поляризации $R_p = 5.6$ Mm и ионизации $R_i = 6.1$ Mm, полученные ниже. Плотность ионов в характерных точках составляет:

$\delta_i(R_{mu}) = 8.4 \cdot 10^{20}$ m$^{-3}$ - нижняя граница мантии;
$\delta_i(R_p) = 7.8 \cdot 10^{16}$ m$^{-3}$ - условная граница поляризации;
$\delta_i(R_i) = 7.1 \cdot 10^{13}$ m$^{-3}$ - условная граница ионизации;
$\delta_i(R_{mo}) = 2.4 \cdot 10^{-6}$ m$^{-3}$ - верхняя граница мантии.

Как видно из рисунка, нижняя часть мантии сильно ионизирована. Плотность ионов падает по экспоненте и в области условной границы ионизации $R_i$ достигает критического значения. Ниже приведена степень ионизации в характерных точках:

$\delta_i(R_{mu})/\Sigma\, m_i(R_{mu}) = 10^{-8}$, $\quad \delta_i(R_p)/\Sigma\, m_i(R_p) = 1.6 \cdot 10^{-12}$ и $\quad \delta_i(R_i)/\Sigma\, m_i(R_i) = 1.8 \cdot 10^{-15}$

где: $\Sigma\, m_i(r)$ - молекулярная плотность смеси (количество молекул в единице объёма). На один ион в среднем приходится не менее $10^8$ нейтральных молекул. Это справедливо и для свободных электронов, парциальным давлением которых в сравнении с парциальным давлением атомов в модели идеального газа можно пренебречь. Таким образом, вещество мантии находится в состоянии электронно-дырочной плазмы, ионы которой „вморожены" в основное вещество, а свободные электроны находятся в состоянии электронного газа. Другими словами, возникают условия для градиентной поляризации свободных электронов в мантии Земли.

## ПОЛЯРИЗАЦИЯ ЗАРЯДОВ В МАНТИИ ЗЕМЛИ

Под действием радиальной составляющей градиентов температуры и плотности свободные электроны смещаются в направлении поверхности Земли (поляризуются), и плазма переходит в новое равновесное состояние, описываемое уравнением равновесия сил, действующих на элементарный объём электронного газа в электрическом поле, создаваемом нескомпенсированным положительным зарядом ионов:

$$E_{pi}(r) \cdot \Delta q_e(r) + S_n(r) \cdot \Delta p(r) = 0 \qquad (2)$$

где: $E_{pi}(r)$ - напряженность электростатического поля нескомпенсированного положительного заряда ионов, $\Delta q_e(r)$ - распределённый в элементарном объёме заряд свободных электронов, $S_n(r)$ - перпендикулярная радиусу площадь элементарного объёма, $\Delta p(r)$ - разность давлений на противоположных поверхностях элементарного объёма в радиальном направлении.

Интегрируя (2), получим распределение нескомпенсированного положительного заряда ионов мантии, который при отрицательных значениях функции градиентов имеет реальное значение и характеризует смещение свободных электронов от центра к поверхности, а при мнимом значении - смещение к центру Земли.

$$q_p(r) = \sqrt{-2\varepsilon_o \int_{R_{mu}}^{r} \varepsilon(r) \cdot S_n(r)^2 \cdot p(r) \cdot \psi(r) \cdot dr} \qquad (3)$$

где: $\varepsilon_o$ - электрическая постоянная; $\varepsilon(r)$ - относительная диэлектрическая проницаемость мантии; $S_n(r) = 4\pi r^2$ - площадь сферической поверхности; $p(r) = n(r) \cdot k_b \cdot T_z(r)$ - давление идеального газа, $n(r) = \Sigma\, na_i(r)$ - количество атомов в единице объёма (окружающая свободный электрон среда состоит из более $10^8$ нейтральных атомов), $na_i(r) = N_i \cdot m_i(r)$ - плотность атомов $i$ - го окисла, $N_i$ - число атомов в молекуле окисла (табл.2), $m_i(r)$ - плотность молекул окисла, подробнее в (1), $T_z(r)$ – температура мантии; $\psi(r) = (\nabla_r n(r)/n(r) + \nabla_r T_z(r)/T_z(r))$ - функция градиентов (функция, равная сумме относительных значений радиальных градиентов плотности атомов и температуры мантии.

Ниже $R_{mu}$ расположено проводящее жидкое ядро и поляризация зарядов в рамках рассматриваемого механизма невозможна.

Уравнение (2) теряет физический смысл при низкой плотности свободных электронов. Каждый атом должен испытать в среднем за секунду не менее одного соударения с распределёнными в объёме свободными электронами:

$$N_e(r) = u_s(r) \cdot \sqrt[3]{n(r)} \cdot kp_{e/a}(r) \geq 1/s \qquad (4.1)$$

где: $u_s(r) = \sqrt{8 \cdot k_b \cdot T_z(r) / \pi \cdot m_e}$ - средняя арифметическая скорость движения свободных электронов [10, с.207], $m_e$ - масса электрона, $n(r)$ - плотности атомов (количество атомов в единице объёма), $kp_{e/a}(r) = \delta_i(r)/n(r)$ - отношение плотности свободных электронов к плотности атомов.

Неравенство (4.1) является физическим условием применения выражения (3). Его решение даёт значение границы ионизации $R_i$ = 6.1 Mm, рис.3.

В качестве критерия границы поляризации свободных электронов рассмотрим следующее выражение, график которого представлен на рис.3:

$$\ln(N_D(r) \cdot k_p(r)) \leq 0 \qquad (4.2)$$

где: $N_D(r) = (4/3) \cdot \pi \cdot R_D(r)^3 \cdot \delta_i(r)$ - Дебаевское число плазмы (число частиц заряда одного знака в пределах области, ограниченной $R_D(r) = (\varepsilon_0 \cdot \varepsilon(r) \cdot k_b \cdot T_Z(r) / \delta_i(r) \cdot q_e^2)^{1/2}$ - Дебаевским радиусом экранирования); $k_p = \delta_p(r)/\delta_i(r)$ - коэффициент поляризации, $\delta_p(r) = dq_p(r)/(q_e \cdot dV)$ - плотность «нескомпенсированных» ионов (отношение плотности нескомпенсированного в мантии Земли положительного заряда к элементарному заряду), $dV = S_n(r) \cdot dr$ - элементарный объём, $dq_p(r)/dr = -\varepsilon_0 \cdot \varepsilon(r) \cdot S_n(r)^2 \cdot p(r) \cdot \psi(r)/q_p(r)$ - частная производная от заряда по радиусу, полученная из (3), $\delta_i(r)$ - плотность ионов (1).

С физической точки зрения выражение (4.2) означает, что среднее число свободных электронов, которые в результате поляризации «покидают» область, ограниченную Дебаевским радиусом экранирования, не превышает единицы и плазма остаётся квазинейтральной.

Так как разделение зарядов в теле Земли невозможно [1, Т2, с.82], решение выражения (4.2) даёт величину границы поляризации $R_p$ = 5.6 Mm < $R_i$ = 6.1 Mm, значение которой расположено в области выраженной ионизации, рис.3.

Относительная диэлектрическая проницаемость среды для однородной смеси не зависит от радиуса:

$$\varepsilon(r) = \varepsilon = \sum_i (mt_i \cdot \varepsilon_i) / \sum_i mt_i = 12.5$$

где: $mt_i = nt_i / M_i$ - доля молекул *i*-го окисла, $M_i$ - молярная масса молекулы *i*-го окисла, $nt_i = pr_i / \Sigma\, pr_i$ - коэффициент содержания окисла в смеси, $pr_i$ - массовая доля *i*-го окисла в процентах, $\varepsilon_i$ - относительная диэлектрическая проницаемость *i*-го окисла, табл.2.

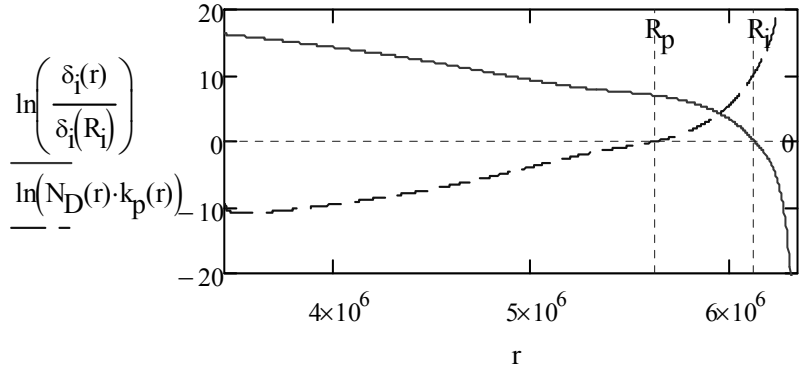

**Рис.3.** Ионизация мантии и критерий поляризации.

В дальнейшем обозначение относительной диэлектрической проницаемости среды: $\varepsilon$ - в варианте однородной смеси и $\varepsilon(r)$ - в структурных вариантах, рассмотренных ниже. Выражение (3) используется для расчёта магнитного момента и интеграла сил, действующих на свободные электроны за пределами границы поляризации.

**Распределение свободных электронов за границей поляризации**

Распределение свободных электронов в потенциальном электростатическом поле подчиняется закону распределения Больцмана [10, с.224]:
$$n_e(r) = n_o \cdot e^{-W(r)/k_b T_z(R_p + r)} \tag{5.1}$$

где: $n_o = q_p(R_p)/q_e$ - число свободных электронов за границей поляризации (в результате поляризации нескомпенсированный заряд свободных электронов равен нескомпенсированному положительному заряду ионов), $q_e$ - элементарный заряд, $q_p(R_p)$ - нескомпенсированный заряд ионов на границе поляризации, $T_z(R_p+r)$ - температура мантии вблизи границы поляризации, $W(r)$ - потенциальная энергия электрона в электростатическом поле.

Взяв производную по объёму от обеих частей уравнения (5.1) и заменив выражение $W(r)/k_b T_z(R_p+r)$ его дифференциалом в точке нулевого потенциала, получим линейное приближение плотности распределения свободных электронов за границей поляризации:
$$\rho_e(r) = \rho_o \cdot e^{-r/\delta R}$$

где: $\rho_o = n_o/(S_n(R_p)\,\delta R) = 1.5 \cdot 10^{29}$ m$^{-3}$ - плотность свободных электронов в точке нулевого потенциала, $\delta R = k_b T_z(R_p)/W'_r(0) = 34$ pm имеет размерность длины и характеризует расстояние, на котором плотность электронов уменьшается в $e$ раз ($e$ - основание натурального логарифма).

$$W'_r(0) = F_q(0) = \frac{q_e q_p(R_p)}{\varepsilon \cdot \varepsilon_0 S_n(R_p)}$$

где: $F_q(0)$ – сила, действующая на электрон, в точке нулевого потенциала (принята граница поляризации $R_p$), $S_n(R_p)$ - площадь сферической поверхности, $\varepsilon = 12.5$ - относительная диэлектрическая проницаемость среды.

Толщина слоя, в котором распределены электроны, не превышает **10·$\delta R$~0.34 nm**. Согласно температурному критерию вырождения газов [10, с.229], $T_0 = 1.6 \cdot 10^5$ K $\gg$ $T_z(R_p) = 3 \cdot 10^3$ K, электронный газ подчиняется квантовой статистике Ферми-Дирака.

Элементарная сила, действующая за пределами границы поляризации на нескомпенсированный заряд свободных электронов, находящихся в элементарном объёме, имеет вид:

$$\Delta F(r) = E(r) \cdot \Delta q_{pe} \qquad (5.2)$$

где: $E(r) = (q_p(R_p) + q_{pe}(r))/\varepsilon \cdot \varepsilon_o \cdot S_n(R_p+r)$ - напряжённость электростатического поля, $(q_p(R_p) + q_{pe}(r))$ - суммарный нескомпенсированный заряд за границей поляризации, $q_{pe}(r)$ - нескомпенсированный заряд свободных электронов, $\Delta q_{pe}$ - нескомпенсированный заряд свободных электронов в элементарном объёме.

Интегрируя (5.2) с учётом равенства нескомпенсированных зарядов разного знака, получим интеграл сил притяжения, направленных к центру (знак минус) и действующих на нескомпенсированный заряд свободных электронов за пределами границы поляризации:

$$F_g = \frac{-q_p(R_p)^2}{2\varepsilon_o \varepsilon(R_p) \cdot S_n(R_p)} = -1.3 \text{ YN} \qquad (6)$$

Интеграл сил градиентного давления, направленных от центра и действующих на свободные электроны, распределённые в мантии Земли:

$$F_m = -\int_{R_{mu}}^{R_P} S_n(r) \cdot p(r) \cdot \psi(r) \cdot dr = 2.1 \text{ YN} \qquad (7)$$

Это значение на 39% превышает интеграл сил притяжения и подтверждает возможность удержания зарядов в поляризованном состоянии.

## СРАВНЕНИЕ ПОЛУЧЕННОГО РЕЗУЛЬТАТА С ДАННЫМИ ИЗМЕРЕНИЙ

Дипольная мода магнитного момента Земли на 1995 г. Составляла **$M_{1995} = -7.8 \cdot 10^{22}$ A·m$^2$ = –78 ZA·m$^2$** (http://ru.wikipedia.org/). Знак минус указывает на то, что магнитный момент направлен в сторону, противоположную её механическому моменту. Отсюда следует, что северный магнитный полюс Земли совпадает с южным географическим полюсом.

Выражение элементарного магнитного момента, возникающего при вращении заряженного шара, имеет вид: $\Delta M_d = S_d \cdot \Delta I$

где: $S_d = \pi \cdot (r \cdot \cos(\theta))^2$ - площадь, ограниченная контуром тока; $\Delta I = v \cdot \delta_p(r) \cdot \Delta V$ - элементарный контур тока, $v$ - частота вращения, $q_e$ - заряд электрона, $\delta_p(r)$ - плотность нескомпенсированного заряда ионов в мантии; $\Delta V = 2\pi r \cdot \cos(\theta) \cdot r \cdot \Delta\theta \cdot \Delta r$ - элементарный объём, $\theta$ – широта, $r$ - радиус удаления от центра Земли.

Так как предлагаемая модель рассматривает только поляризацию зарядов, то суммарный заряд за пределами границы поляризации равен нулю. Выражение магнитного момента, создаваемого вращением поляризованного шара (назовём его ротационным магнитным диполем, чтобы отличать от дипольной моды магнитного поля Земли, наклонённой к оси её вращения под углом 11°), после ряда преобразований и интегрирования по частям имеет вид:

$$M_d = \frac{-2\Omega_z}{3} \cdot \int_{R_{mu}}^{R_P} r \cdot q_p(r) \cdot dr \qquad (8)$$

где: $\Omega_z = 7.3 \cdot 10^{-5}$ s$^{-1}$ - угловая скорость вращения Земли, $q_p(r)$ - нескомпенсированный положительный заряд ионов в мантии (3).

Поскольку поляризация зарядов в ядре невозможна, нижняя граница интегрирования соответствует нижней границе мантии $R_{mu}$, а верхняя, - границе поляризации $R_p$. Значение ротационного магнитного момента в варианте однородной смеси составило
**$M_{1d} = -110$ ZA·m$^2$ = 1.4·$M_{1995}$**.

Полученное в результате интегрирования по частям отрицательное значение свидетельствует о том, что вектор ротационного магнитного момента направлен противоположно вектору механического момента вращения шара.

## ВАРИАНТЫ СТРУКТУРНОГО РАСПРЕДЕЛЕНИЯ ОКИСЛОВ

В связи с малым процентным содержанием CaO в мантии наиболее вероятно, что он представлен комплексным соединением CaMg(Si$_2$O$_6$) - диопсид (diopside). Массовые доли SiO$_2$ и MgO в диопсиде, исходя из пропорций к массовой доле CaO (табл.2), составляют $m_{SiO2} = 6.9\%$ и $m_{MgO} = 2.3\%$ в массовой доле мантии. Ввиду малого процентного содержания Al$_2$O$_3$ наиболее вероятно, что он также представлен лишь комплексными соединениями.

### Границы распределения окислов и связанные с ними параметры

В рассматриваемом варианте окислы полностью разделены и упорядочены по возрастанию их плотности. Зная процентное содержание окислов в мантии (табл.2), легко получить границы их распределения. Зная границы, можно получить центры распределения основных веществ мантии и плотности атомов в этих точках, расчётные значения которых приведены в табл.3. Граница между MgO и Al$_2$O$_3$ вероятнее всего состоит из MgO(Al$_2$O$_3$), а между Al$_2$O$_3$ и FeO - из FeO(Al$_2$O$_3$) - шпинелей (spinel), расчётное значение диэлектрической проницаемости $\varepsilon_{FeO(Al2O3)} = 55$. Граница между MgO(Al$_2$O$_3$) и FeO(Al$_2$O$_3$) пролегает по центру распределения Al$_2$O$_3$. Плотность атомов в этой точке рассчитана для комплексного соединения [FeO(50%)+MgO(50%)]Al$_2$O$_3$.

**Табл.3.** Распределение окислов основных элементов в мантии Земли.

| Вещество | Al$_2$O$_3$ | MgO | CaMg[Si$_2$O$_6$] | SiO$_2$ |
|---|---|---|---|---|
| Нижняя граница $g$, Mm | 3.7 | 3.9 | 5.05 | 5.4 |
| Центр распределения ($Xp$), Mm | 3.8 | 4.5 | 5.2 | 5.9 |
| Плотность атомов ($Na_2$), m$^{-3}$ | 1.7·10$^{29}$ | 1.5·10$^{29}$ | 1.3·10$^{29}$ | 1.1·10$^{29}$ |

На основе полученных данных (табл.3) выполнена интерполяция параболическими сплайнами плотности атомов $n_2(r)$ и линейная интерполяция относительной диэлектрической проницаемости вещества $\varepsilon_2(r)$ на основе данных табл.2. Результаты интерполяций приведены на рис.4.

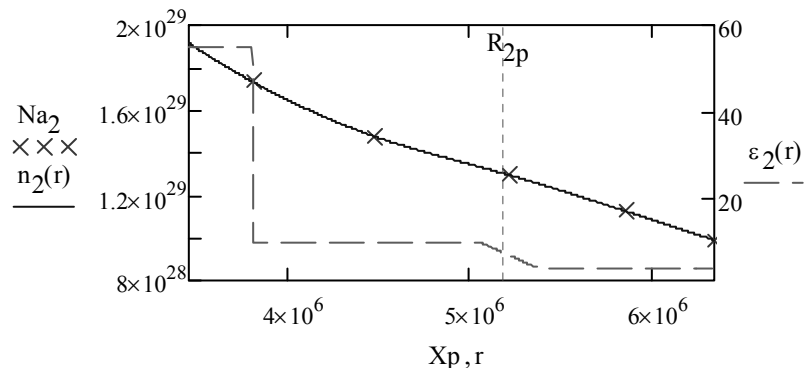

**Рис.4.** Плотность атомов $n_2(r)$ и относительная диэлектрическая проницаемость $\varepsilon_2(r)$ мантии.

Вертикальным маркером отмечено новое значение границы поляризации, полученное ниже. Подстановка зависимостей $n_2(r)$ и $\varepsilon_2(r)$ в (3), (6) и (7) даёт возможность вычислить значения заряда, интегралов сил и магнитного момента (8) с учетом изменения плотности атомов вещества и диэлектрической проницаемости в мантии Земли.

### Границы интегрирования второго варианта и функции градиентов

Для уточнения верхней границы интегрирования обратим внимание на то, что в этом варианте существуют области, в которых выполняется равенство $F_m(r) + F_g(r) = 0$ интегралов сил градиентного давления (7) и сил притяжения (6). Это равенство достигается в зонах изменения диэлектрической проницаемости в точках $R_2 = 3.8$ Mm и $R_{2p} = 5.3$ Mm, удовлетворяющих первому условию ($R_2 < R_{2p} < R_{1p} = R_p$).

На рис.5. представлены зависимости функций градиентов (индексный номер соответствует номеру варианта), с указанием вертикальными маркерами верхних границ интегрирования первого - $R_{1p}$ и второго - $R_{2p}$ вариантов. Наименьшее значение радиуса соответствует нижней границе мантии $R_{mu}$.

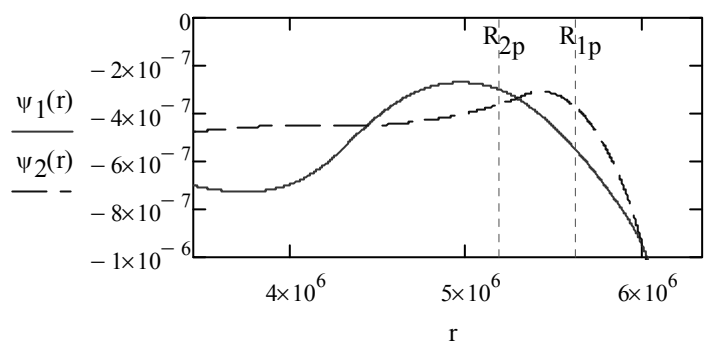

**Рис.5.** Функции градиентов двух вариантов.

Эти зависимости использованы в (3) при расчёте зарядов. Значение интегралов сил (7) и (6) $F_m = -F_g = 1.1$ YN при относительной погрешности результата вычислений ($F_m + F_g$)/$F_m$ = 0.01%. Расчётное значение магнитного момента (8) составило: $M_{2d} = -$**54 ZA·m$^2$ = 0.7·$M_{1995}$**.

### Комбинированный вариант распределения окислов

В этом варианте окислы в верхней мантии, - $SiO_2$ и комплексное соединение $CaMg(Si_2O_6)$, дифференцированы по типу второго варианта, а в нижней мантии по типу первого варианта образуют однородную смесь из $FeO$, $Al_2O_3$ и $MgO$, расчётное значение относительной диэлектрической проницаемости которой составило $\varepsilon_3 = 18$.

### ОБСУЖДЕНИЕ РЕЗУЛЬТАТА

На рис.6 представлено распределение нескомпенсированного заряда в мантии Земли для трёх вариантов, использованное в (8) при расчете магнитного момента. Максимальное значение заряда первого ($q_{1m} = 0.33$ PC) второго ($q_{2m} = 0.22$ PC) и третьего ($q_{3m} = 0.35$ PC) вариантов достигается на границе поляризации. Для среднего арифметического значения тока $I_{3s} = 2.9$ GA, создаваемого средним арифметическим значением заряда $q_{3s} = 0.25$ PC рассчитаны условные радиусы (средние арифметические значения) протекания токов положительных ($R_{3i} = 3.9$ Mm) и отрицательных ($R_{3e} = 5.04$ Mm) зарядов в мантии Земли, соответствующие третьему варианту (отмечены вертикальными маркерами). Величина границы поляризации в этом варианте $R_{3p} = 5.2$ **Mm.** Расчётное значение магнитного момента (8) составило: $M_{3d} = -94$ ZA·m$^2$ = 1.2·$M_{1995}$. Относительная разница интеграла сил градиентного давления (7) на распределённые в мантии Зем-ли свободные электроны и интеграла сил притяжения нескомпенсированных свободных электронов (6) на границе поляризации: $(F_{3m} + F_{3g})/F_{3m} = 0.01\%$, что характеризует состояние как равновесное. Значение интегралов сил: $F_{3m} = -F_{3g} = 1.7$ **YN**.

Среднее взвешенное значение тока находится в пределах между $I_{3s}$ = **2.9 GA** и $I_3(R_{3e})$ = **3.9 GA**, $R_{3e}$ = **5.04 Mm**. Полученный результат хорошо согласуется с оценочными расчетами Дж. Орира, согласно которым современную напряженность магнитного поля Земли может обеспечить кольцевой электрический ток силой $3.4 \cdot 10^9$ **A** = **3.4 GA**, протекающий в плоскости экватора на расстоянии **5000 km = 5 Mm** от центра планеты [13].

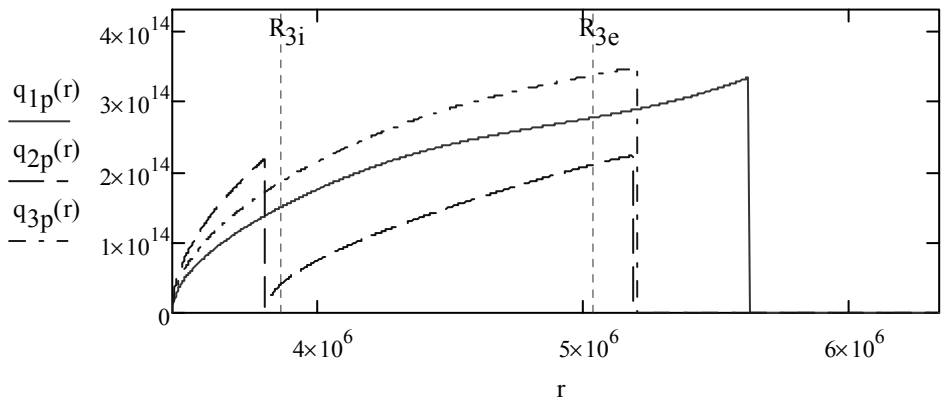

**Рис.6.** Распределение нескомпенсированного заряда в мантии Земли.

Схематично ротационный магнитный диполь можно представить в виде разности магнитных диполей, создаваемых двумя кольцевыми токами величиной $I_{3s}$, протекающими в плоскости экватора в разных направлениях на расстоянии от центра $R_{3i}$ (по часовой стрелке) и $R_{3e}$ (против часовой стрелки, если наблюдать со стороны южного географического полюса). Результирующий магнитный момент имеет значение $\pi(R_{3i}^2 - R_{3e}^2)I_{3s}$ = **–94 ZA·m²** и направлен в сторону южного географического полюса Земли.

Предложенный механизм даёт хорошее описание величины дипольной моды магнитного поля Земли. Выполненные для трёх вариантов расчеты убедительно показывают возможность поляризации зарядов в её мантии. Все варианты являются виртуальными, а первые два отражают крайние возможные состояния вещества мантии. Реальное значение величины ротационного магнитного диполя может находиться в диапазоне **(0.7 ÷ 1.4)·$M_{1995}$**. Его ожидаемое значение должно превышать величину магнитного диполя Земли примерно на 10%, так как на дипольную моду приходится около 90% напряженности магнитного поля [8, с.266]. Комбинированный вариант распределения окислов даёт близкое к этой величине значение, превышая его примерно на 10%. На основании этого можно предположить, что процесс сепарации вещества в нижней мантии находится в начальной фазе.

## ВЫВОДЫ

1) Тепловая ионизация, поляризация свободных электронов под действием градиентов плотности и температуры, скорость углового вращения Земли - основные физические факторы формирования дипольной моды магнитного поля.

2) Магнитный момент Земли формируется за пределами ядра в мантии и, следовательно, жидкая часть ядра пронизана силовыми линиями магнитного поля. Плотность ионов в жидком ядре не менее $\delta_i(R_j)=10^{23}$ **m⁻³**. Конвективные потоки проводящего вещества, возникающие в жидком ядре, могут вызывать локальные изменения магнитного поля у поверхности Земли, а их асимметрия может быть причиной смещения оси магнитного момента. Рассматриваемая как "белый шум" турбулентная составляющая может быть источником инверсии магнитного поля Земли [9]. Другими словами, моды выше дипольной являются результатом рассеяния части энергии ротационного магнитного поля турбулентными потоками, возникающими в жидком ядре, а вектор смещения магнитной оси обусловлен асимметрией конвективных потоков относительно оси вращения Земли.

3) Предложенный механизм не отвергает теории гидромагнитного динамо, а дополняет её, по мнению автора, недостающей стационарной компонентой. Гидромагнитное динамо не является обособленным механизмом генерации магнитного поля и не претендует в этом смысле на полную универсальность [8, с.265].

**Примечание**

Расчеты выполнены с использованием программы MathCAD со стандартной точностью 4 значащие цифры. Значения фундаментальных констант рекомендованы Национальным Институтом Стандартов и Технологии США [2, с.94]. Основные физические параметры Земли соответствуют приведенным в справочнике [1, Т2, с.78-79] значениям.

## СПИСОК ЛИТЕРАТУРЫ

1. Физическая Энциклопедия / гл. редактор А. М. Прохоров. М.: Советская энциклопедия. 1990, в 5-ти томах.
2. Mohr, Peter J. The NIST Reference on Constants, Units, and Uncertainty, from CODATA Recommended Values of the Fundamental Constants - 2006. / By Peter J. Mohr, Barry N. Taylor and David B. Newell // National Institute of Standards and Technology.
3. Монин, А. С. История Земли / А.С.Монин. М.: Наука, 1980 - 224с.
4. Жарков, В. Н. ГЕОФИЗИЧЕСКИЕ ИССЛЕДОВАНИЯ ПЛАНЕТ И СПУТНИКОВ / В. Н. Жарков, // ВЕСТНИК Отделения наук о Земле РАН, № 1(21), 2003г.: Электронный научно-информационный журнал.
(http://www.scgis.ru/russian/cp1251/h_dgggms/CD-R/zharkov/scpub-3_index.html)
5. Справочник химика / гл. редактор Б. П. Никольский. Издание второе, М.-Л.: Химия 1966(Т.1), 1964(Т.2).
6. Иориш, В. С. Термические Константы Веществ. / В. С. Иориш и В. С. Юнгман // Институт теплофизики экстремальных состояний РАН Объединенного института высоких температур РАН, Химический факультет Московского Государственного Университета им. М.В.Ломоносова.: Электронная база данных, рабочая версия - 2 (http://www.chem.msu.ru/cgi-bin/tkv.pl?show=welcome.html).
7. Sharon, G. L./ G. L. Sharon, I. E. Bartmess, J. F. Liebman, J. L. Holmes, R. D. Levin, and W. G. Hallard, //J.Phys. Chem. Ref. Data 17, Suppl. 1 (1988).
8. Зельдович, Я. Б. Гидромагнитное динамо, как источник планетарного, солнечного и галактического магнетизма: / Я. Б. Зельдович, А. А. Рузмайкин //УСПЕХИ ФИЗИЧЕСКИХ НАУК. Том 152, вып.6, Июнь 1987, с.263-284.
9. Pétrélis, F. Simple Mechanism for Reversals of Earth's Magnetic Field / F. Pétrélis, S. Fauve, E. Dormy and J-P. Valet // Physical Review Letters. 102, 144503 (2009).
10. Яворский, Б.М. Справочник по физике / Б.М.Яворский и А.А.Детлаф. М.: Наука, 1968, 940с.
11. Berhanu, M. Magnetic field reversals in an experimental turbulent dynamo /M. Berhanu, R. Monchaux, S. Fauve, N. Mordant, F. Pétrélis, A. Chiffaudel, F. Daviaud, B. Dubrulle, L. Marié, F. Ravelet, M. Bourgoin, Ph. Odier, J.-F. Pinton and R. Volk et al // 2007 EPL 77 59001 (5pp).
12. Долгинов, А.З. О происхождении магнитных полей Земли и небесных тел./ А.З.Долгинов //УСПЕХИ ФИЗИЧЕСКИХ НАУК. Том 152, вып.6, Июнь 1987, с.231-262.

13. Орир Дж. Физика. / Дж. Орир. Пер. с англ. Т.2. - М.: Мир, 1981. - 366 с.

*Статья публикуется впервые.*

*25/05/2011*

**Сведения об авторе**
Стяжкин Олег Владимирович, дипломированный инженер.
Тел. (+49) 030 60974392,   olegstaz@live.ru